\documentclass{ws-ijmpd}
\usepackage[super,compress]{cite}

\usepackage{subfigure}   
\usepackage{graphics}
\usepackage{cite}
\usepackage{hyperref}

\begin{document}

\markboth{Francis Halzen}
{The Observation of High-Energy Neutrinos from the Cosmos}

%
\catchline{}{}{}{}{}
%

\title{The Observation of High-Energy Neutrinos from the Cosmos: Lessons Learned for Multimessenger Astronomy
}

\author{FRANCIS HALZEN
}

\address{Department of Physics, University of Wisconsin–Madison,\\
Madison, WI 53705,
US
}



\maketitle

\begin{history}
\end{history}

\begin{abstract}
The IceCube neutrino telescope discovered PeV-energy neutrinos originating beyond our Galaxy with an energy flux that is comparable to that of GeV-energy gamma rays and EeV-energy cosmic rays. These neutrinos provide the only unobstructed view of the cosmic accelerators that power the highest energy radiation reaching us from the universe. We will review the results from IceCube's first decade of operations, emphasizing the measurement of the diffuse multiflavored neutrino flux from the universe and the identification of the supermassive black hole TXS 0506+056 as a source of cosmic neutrinos and, therefore, cosmic rays. We will speculate on the lessons learned for multimessenger astronomy, among them that extragalactic neutrino sources may be a relatively small subset of the cosmic accelerators observed in high-energy gamma rays and that these may be gamma-ray-obscured at the times that they emit neutrinos.
\end{abstract}




\section{Neutrino Astronomy: a Brief Introduction}
\vspace{.2cm}

The shortest wavelength radiation reaching us from the universe is not radiation at all; it consists of cosmic rays---protons and high-energy nuclei. Some reach us with extreme energies exceeding $10^8$~TeV from a universe beyond our Galaxy that is obscured to gamma rays and from which only neutrinos reach us as astronomical messengers~\cite{Aartsen:2013jdh}. Their origin is still unknown but the identification of a supermassive black hole powering a cosmic-ray accelerator~\cite{IceCube:2018dnn,IceCube:2018cha} represents a breakthrough towards a promising path for resolving the century-old puzzle of the origin of cosmic rays: multimessenger astronomy.

The rationale for searching for cosmic-ray sources by observing neutrinos is straightforward: in relativistic particle flows near neutron stars or black holes, some of the gravitational energy released in the accretion of matter is transformed into the acceleration of protons or heavier nuclei, which subsequently interact with ambient radiation or matter to produce pions and other secondary particles that decay into neutrinos. For instance, when protons interact with intense radiation fields near the source via the photoproduction processes
\begin{equation}
p + \gamma \rightarrow \pi^0 + p
\mbox{ \ and \ }
p + \gamma \rightarrow \pi^+ + n\,,
\label{eq:delta}
\end{equation}
both neutrinos and gamma rays are produced with roughly equal rates; while neutral pions decay into two gamma rays, $\pi^0\to\gamma+\gamma$, the charged pions decay into three high-energy neutrinos ($\nu$) and antineutrinos ($\bar\nu$) via the decay chain $\pi^+\to\mu^++\nu_\mu$ followed by $\mu^+\to e^++\bar\nu_\mu +\nu_e$. The fact that cosmic neutrinos are inevitably accompanied by high-energy photons transforms neutrino astronomy into multimessenger astronomy.

A main challenge of multimessenger astronomy is to separate these photons, which we will refer to as {\em pionic} photons, from photons radiated by electrons that are accelerated along with the cosmic ray protons. Another challenge is to identify the electromagnetic energy associated with the pionic photons because they do not reach our telescopes with their initial energy. They suffer losses in interactions with the extragalactic background light (EBL), predominantly with microwave photons, via the process $\gamma + \gamma_{\it cmb} \rightarrow e^+ + e^-$. Importantly, they may also lose energy in the source. As is the case when constructing a neutrino beam in a particle physics laboratory, neutrinos require a target that transforms the energy of the proton beam into neutrinos. Powerful neutrino sources within reach of IceCube's sensitivity require a dense target that is likely to be opaque to pionic gamma rays. Additionally losing energy in the source, these may reach Earth with MeV energies or below. We will review the accumulating indications that cosmic neutrinos originate in gamma-obscured sources with their associated multimessenger signals emerging below the detection thresholds of high-energy gamma-ray satellites and ground-based TeV gamma-ray telescopes~\cite{Halzen:2019qkf,Kun_2021}.

\section{High-Energy Cosmic Neutrinos}
\vspace{.2cm}

\begin{figure}[t]
  \centering
   \includegraphics[width=1.0\linewidth]{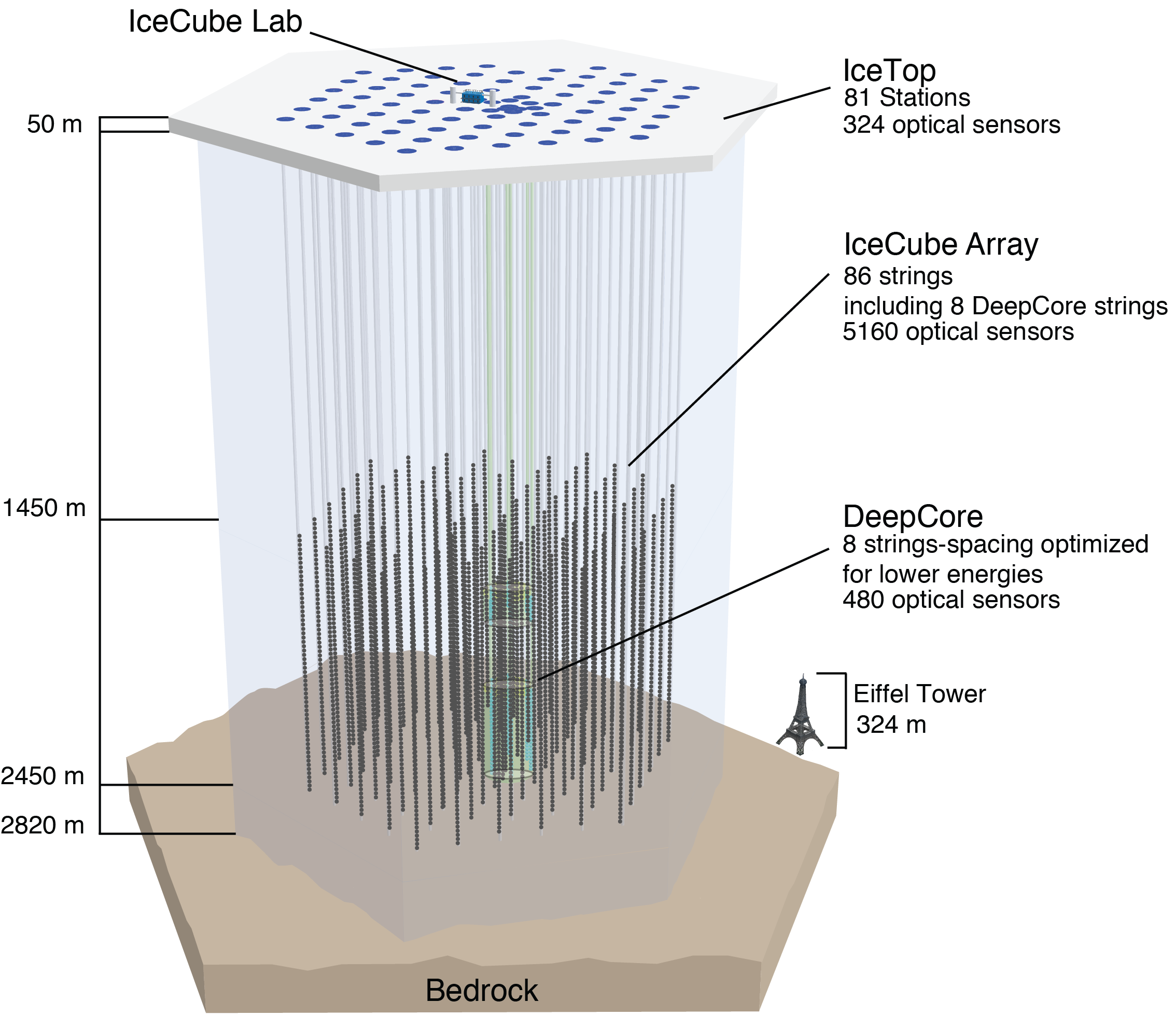}
  \caption{Architecture of the IceCube observatory.}
  \label{fig:array}
\end{figure}

\begin{figure}[t]
  \centering
   \includegraphics[width=0.6\linewidth]{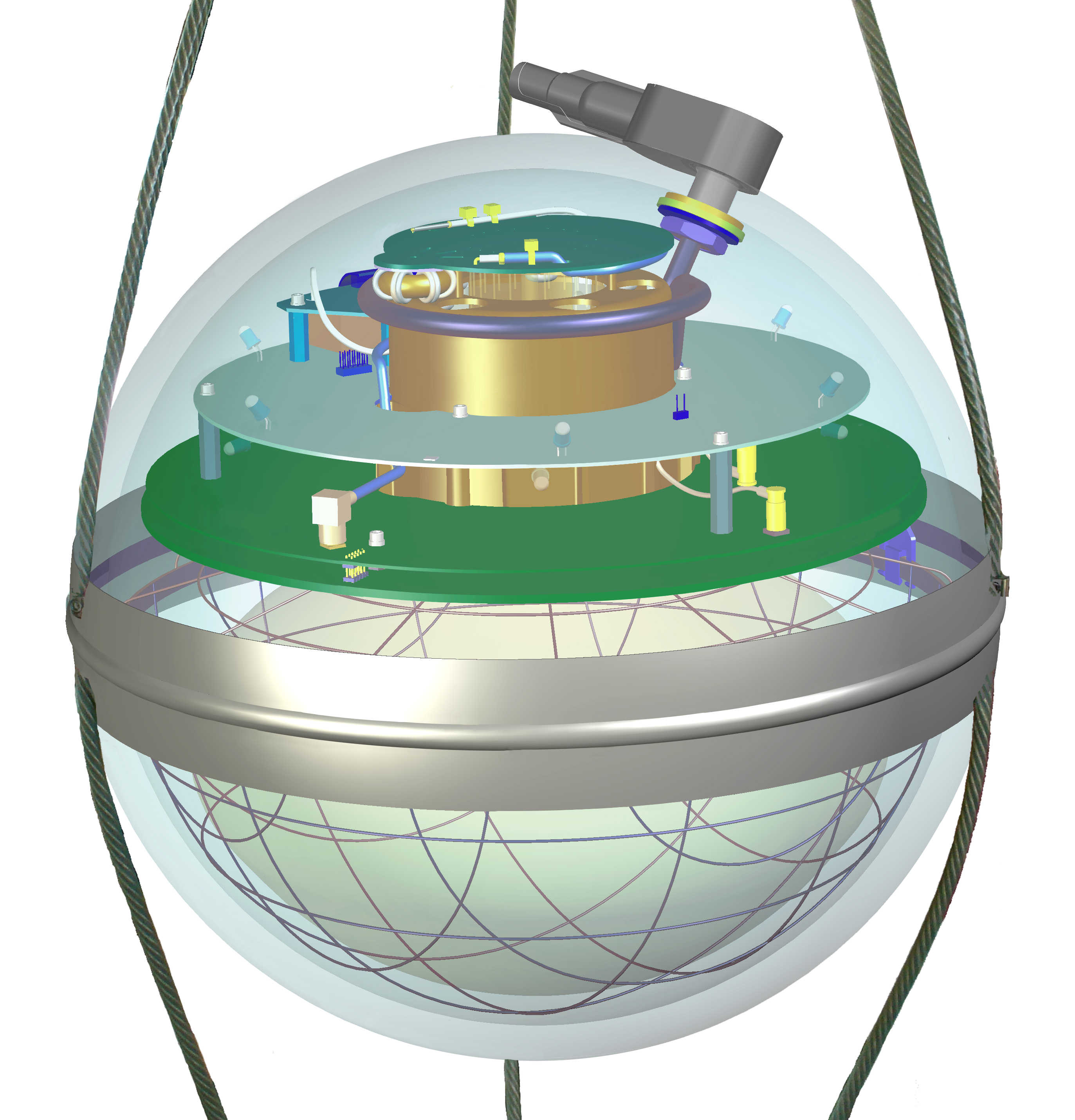}
  \caption{Digital optical module showing the down-facing 10-inch photomultiplier and the associated electronics that digitize the light signals.}
  \label{fig:DOM}
\end{figure}

Close to the National Science Foundation's research station located at the geographical South Pole, the IceCube project~\cite{Aartsen:2016nxy} transformed one cubic kilometer of natural Antarctic ice into a Cherenkov detector. The deep ice of the Antarctic glacier constitutes the detector, forming both support structure and Cherenkov medium. Below a depth of 1,450\,meters, a cubic kilometer of glacial ice has been instrumented with 86 cables called ``strings,'' each of which is equipped with 60~optical sensors; see Fig.~\ref{fig:array}. Each digital optical module (DOM) consists of a glass sphere containing a photomultiplier and the electronics board that captures and digitizes the light signals locally using an onboard computer; see Fig.~\ref{fig:DOM}. The digitized signals are given a global time stamp accurate to 2\,ns and are subsequently transmitted to the surface. Processors at the surface continuously collect the time-stamped signals from the optical modules, each of which functions independently. The digital messages are sent to a string processor and a global event trigger. They are sorted into the Cherenkov radiation patterns that are emitted by muon tracks produced by muon neutrinos interacting in the ice, or by secondary particle showers initiated by electron and tau neutrinos. These reveal the flavor, energy, and direction of the incident neutrino~\cite{Halzen:2006mq}. Constructed between 2004 and 2010, IceCube has now taken 10 years of data with the completed detector.
%


The arrival direction of a secondary muon track, or of an electromagnetic shower initiated by an electron or tau neutrino, is reconstructed from the arrival times of the Cherenkov photons at the optical sensors, while the number of photons is a proxy for the energy deposited by secondary particles in the detector. Although the detector only records the energy of the secondary muon inside the detector, from Standard Model physics we can infer the energy spectrum of the parent neutrino. 

\begin{figure}
\centering
\begin{subfigure}
\centering
\includegraphics[width=.75\textwidth]{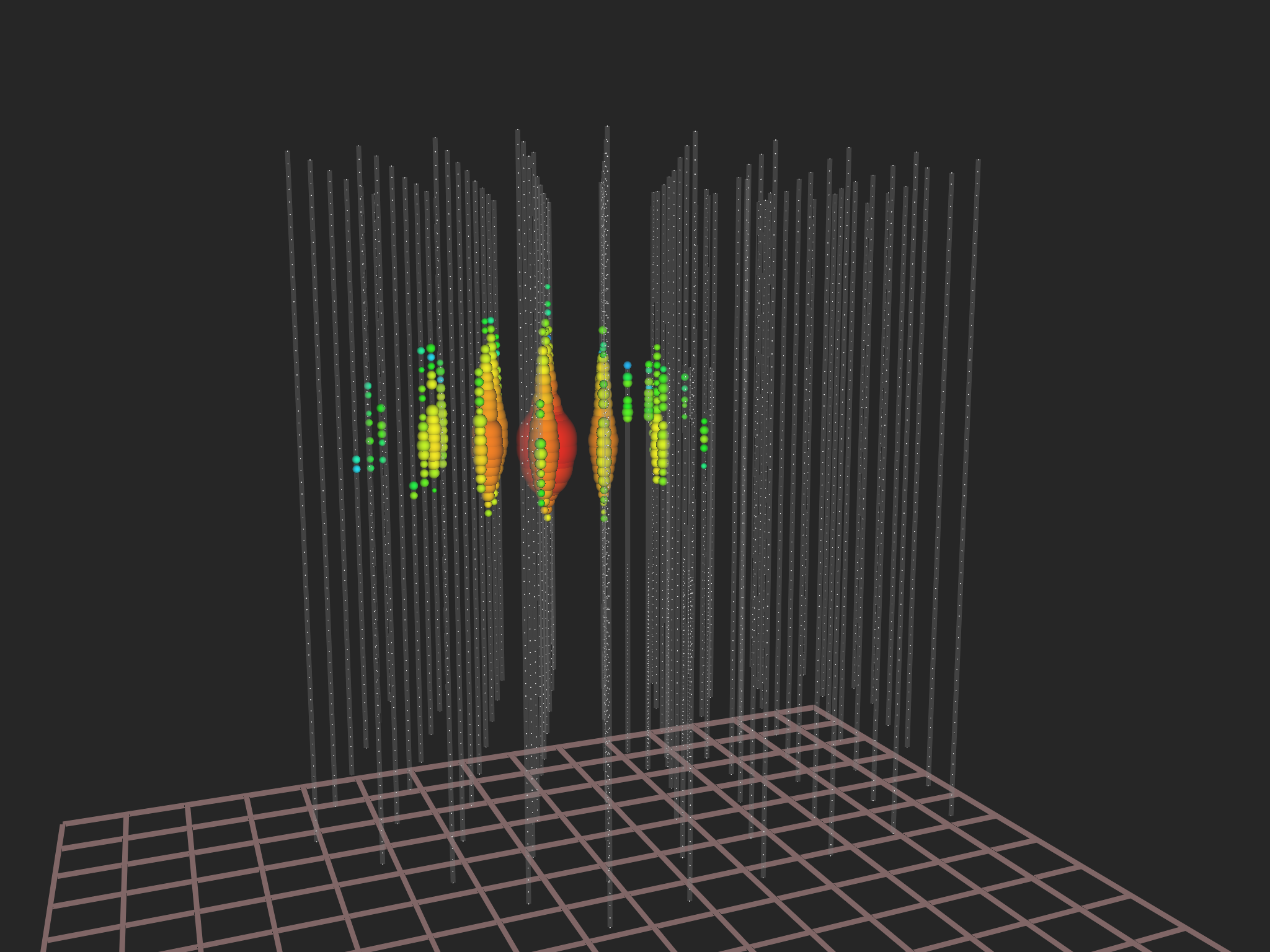}
\end{subfigure}
\begin{subfigure}
\centering
\includegraphics[width=.75\textwidth]{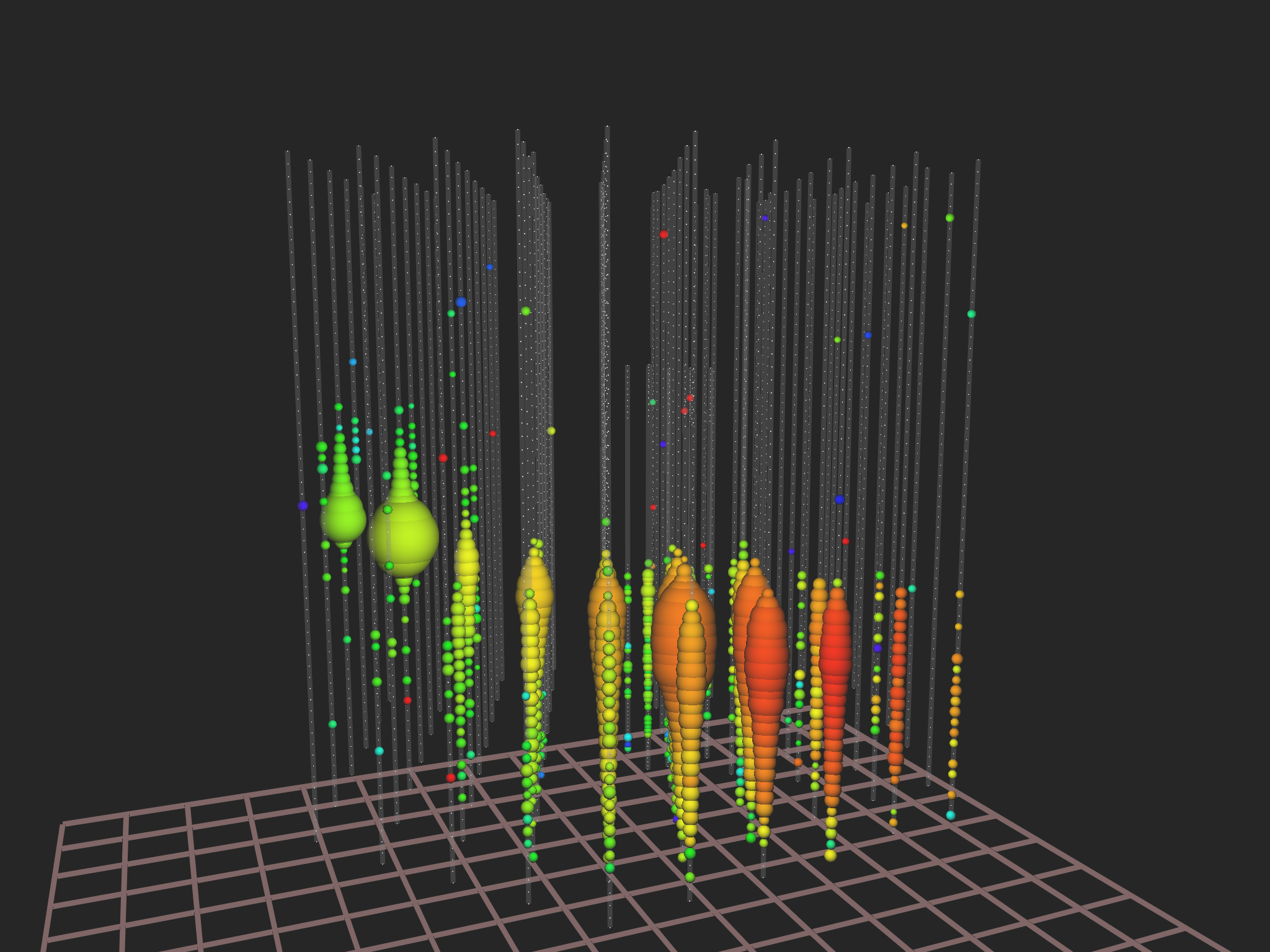}
\end{subfigure}
\caption{{\bf Top Panel:}  Light pool produced in IceCube by a shower initiated by an electron or tau neutrino of $1.14$ PeV, which represents a lower limit on the energy of the neutrino that initiated the shower. White dots represent sensors with no signal. For the colored dots, the color indicates arrival time, from red (early) to purple (late) following the rainbow, and size reflects the number of photons detected. {\bf Bottom Panel:}  A muon track coming up through the Earth, traverses the detector at an angle of $11^\circ$ below the horizon. The deposited energy, i.e., the energy equivalent of the total Cherenkov light of all charged secondary particles inside the detector, is 2.6\,PeV.}
\label{fig:erniekloppo}
\end{figure}

Tracks resulting from muon neutrino interactions can be pointed back to their sources with a $\le 0.4^\circ$ angular resolution for the highest energy events. In contrast, the reconstruction of cascade directions, in principle possible to within a few degrees, is still in the development stage in IceCube, achieving $8^\circ$ resolution~\cite{Aartsen:2013vja, Tyuan2017}. On the other hand, determining the shower energy from the observed light pool is straightforward, and a resolution of better than 15\% can be achieved. For illustration, we contrast in Fig.~\ref{fig:erniekloppo} the Cherenkov patterns initiated by an electron (or tau) neutrino of 1\,PeV energy (top) and a neutrino-induced muon losing 2.6\,PeV energy while traversing the detector (bottom).

IceCube identifies cosmic neutrinos in a large background of muons and neutrinos produced by cosmic rays interacting in the atmosphere. Two principal methods are used to separate neutrinos of cosmic origin from the background of atmospheric muons and neutrinos. The first method reconstructs muon tracks reaching the detector from directions below the horizon, the second identifies neutrinos of all flavors that interact inside the instrumented volume of the detector; examples of events are shown in Fig.~\ref{fig:erniekloppo}. We will describe these methods in turn.

\subsection{Muon Neutrinos Through the Earth}

\begin{figure}[ht]\centering
\includegraphics[width=0.9\columnwidth]{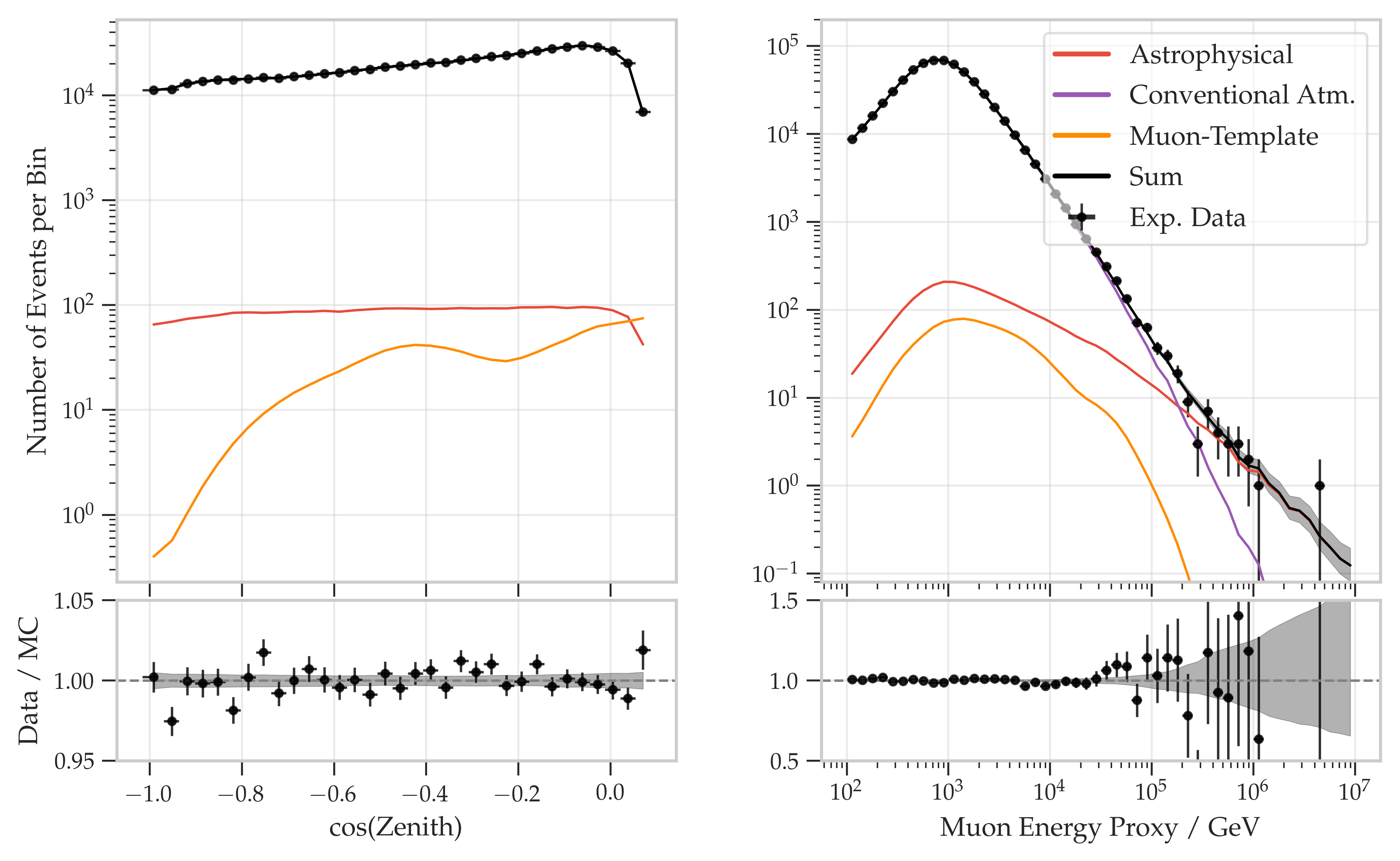}
\caption{The distributions of muon tracks arriving from the Northern Hemisphere, i.e., with declination greater than $-5^\circ$, for the period 2010-18~\cite{Stettner:2019tok}, are shown as a function of reconstructed zenith (left) and muon energy (right). The full data set consists of about 650,000 neutrino events with a purity of 99.7\%. Best fits to the low-energy atmospheric and high-energy astrophysical components of the neutrino flux are superimposed. Statistical errors are shown as crosses, the grey bands in the ratio plots show an estimate of the systematic error obtained by varying all fit-parameters within their uncertainties.}
\label{fig:diffusenumu}
\end{figure}

Detecting particles from directions below the horizon has the immediate advantage of eliminating the overwhelming background of cosmic-ray muons that reach the detector from above. The tracks are separated from the background of atmospheric muons using the Earth as a filter. IceCube thus collects samples of muon neutrinos with high purity, typically well above 99\%, at a rate of more than 100,000 per year. The neutrino energies cover more than six orders of magnitude, from $\sim 5$\,GeV in the highly instrumented inner core, labeled DeepCore in Fig.~\ref{fig:array}, to extreme energies beyond 10~PeV. IceCube thus measured the atmospheric neutrino flux over more than five orders of magnitude in energy with a result that is consistent with theoretical calculations. Muon neutrinos can be detected even when interacting outside the detector because of the kilometer range of the secondary muons. More importantly, IceCube has observed an excess of neutrino events at energies beyond 100\,TeV~\cite{Aartsen:2015rwa,Aartsen:2016xlq,Aartsen:2017mau} that cannot be accounted for by the atmospheric flux at the $5.6 \sigma$ level. A recent measurement of the energy flux covering 9.5 years of data was performed yielding a sample of 650,000 neutrinos with 99.7\% purity; see Fig.~\ref{fig:diffusenumu}. The excess cosmic neutrino flux (red) over the atmospheric background (purple) is well described by a power law with a spectral index of $-2.37\pm0.09$ and a normalization at 100\,TeV neutrino energy of $(1.36^{+0.24}_{-0.25})\,\times10^{-18}\,\rm GeV^{-1}\rm cm^{-2} \rm sr^{-1} \rm s^{-1}$~\cite{Stettner:2019tok}. The residual atmospheric muon background is small (yellow). For more details, see Ref.~\citen{Stettner:811376}.

\begin{figure}
\centering
\begin{subfigure}
\centering
\includegraphics[width=1.0\textwidth]{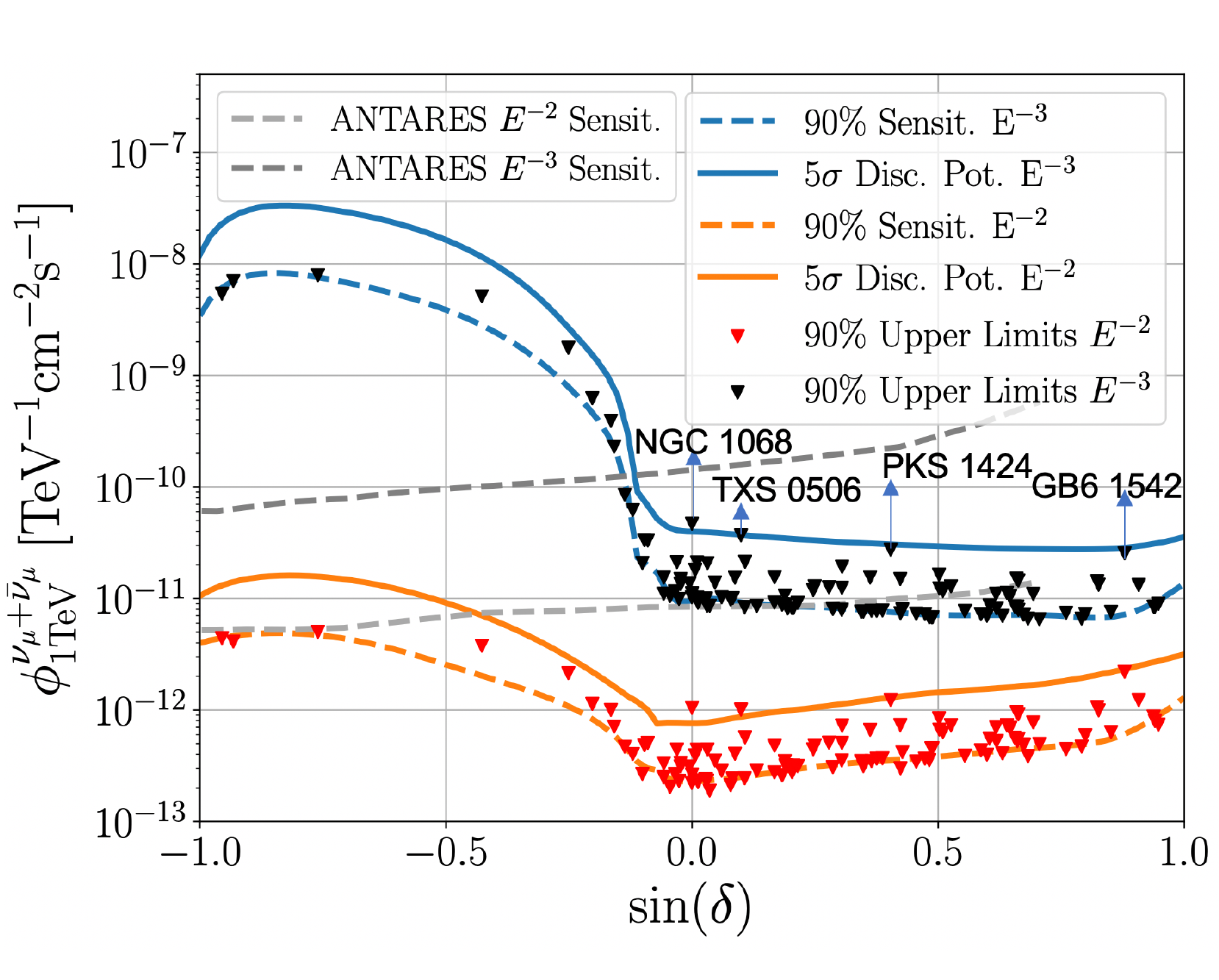}
\end{subfigure}
\begin{subfigure}
\centering
\includegraphics[width=.5\textwidth]{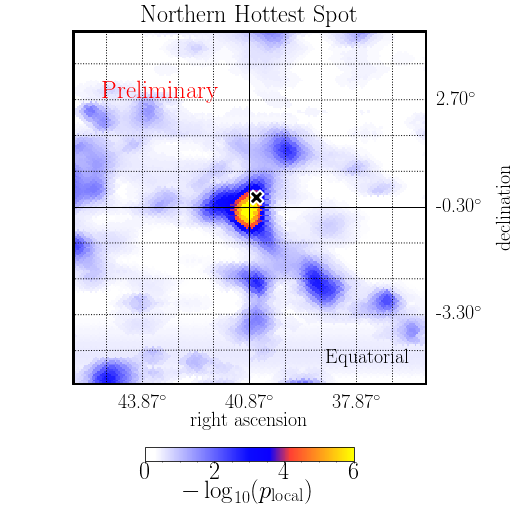}
\end{subfigure}
\caption{{\bf Top Panel:} Upper limits on the flux from candidate point sources of neutrinos in 10 years of IceCube data assuming two spectral indices of the flux~\cite{Aartsen:2019fau}. Also shown as triangles are limits on a preselected list of about one hundred candidate sources. Four sources exceed the $4\sigma$ level (pretrial) and collectively result in a $3\sigma$ anisotropy of the sky map. {\bf Bottom Panel:} Association of the hottest source in the sky map as well as in the list of preselected candidate sources with the active galaxy NGC 1068.}
\label{fig:10year}
\end{figure}

The measured arrival directions of the astrophysical muon tracks are isotropically distributed over the sky. Surprisingly, there is no evidence for a correlation to nearby sources in the Galactic plane; IceCube observes a diffuse flux of extragalactic sources. Only after collecting 10 years of data~\cite{Aartsen:2019fau} did the first evidence emerge at the $3\sigma$ level that the neutrino sky is not isotropic. The anisotropy results from four sources—TXS 0506+056 among them (more about that source later)—that emerge as point sources of neutrinos with a p-value of less than 0.01 (pretrial); see Fig.~\ref{fig:10year}. The strongest of these sources is the nearby active galaxy NGC 1068, also known as Messier 77, which, independently, also emerges as the most significant source in a list of about a hundred candidates that had been preselected by the collaboration as targets of interest.

\subsection{Neutrinos Interacting Inside the Instrumented Volume}

The second method for separating cosmic from atmospheric neutrinos exclusively identifies high-energy neutrinos interacting inside the instrumented volume of the detector, so-called starting events. After only two years of operation, IceCube used this method to make the initial discovery of an extragalactic flux of cosmic neutrinos~\cite{Aartsen:2013jdh} with an energy flux, $E^2 dN/dE$, in the local universe that is, surprisingly, similar to that in gamma rays~\cite{Ackermann:2014usa,Fang:2017zjf}. Using this method, one divides the instrumented volume of ice into an outer veto shield and a $\sim500$-megaton inner fiducial volume. The advantage of focusing on neutrinos interacting inside the instrumented volume of ice is that the detector functions as a total absorption calorimeter~\cite{Aartsen:2013vja}, allowing for a good energy measurement that separates cosmic from lower-energy atmospheric neutrinos. In contrast to the first method, neutrinos from all directions in the sky and of all flavors can be identified, including both muon tracks and secondary showers produced by charged-current interactions of electron and tau neutrinos and neutral current interactions of neutrinos of all flavors. A sample event with a light pool of roughly one hundred thousand photoelectrons extending over more than 500 meters is shown in the top panel of Fig.~\ref{fig:erniekloppo}.

\begin{figure}[t]\centering
\includegraphics[width=\textwidth]{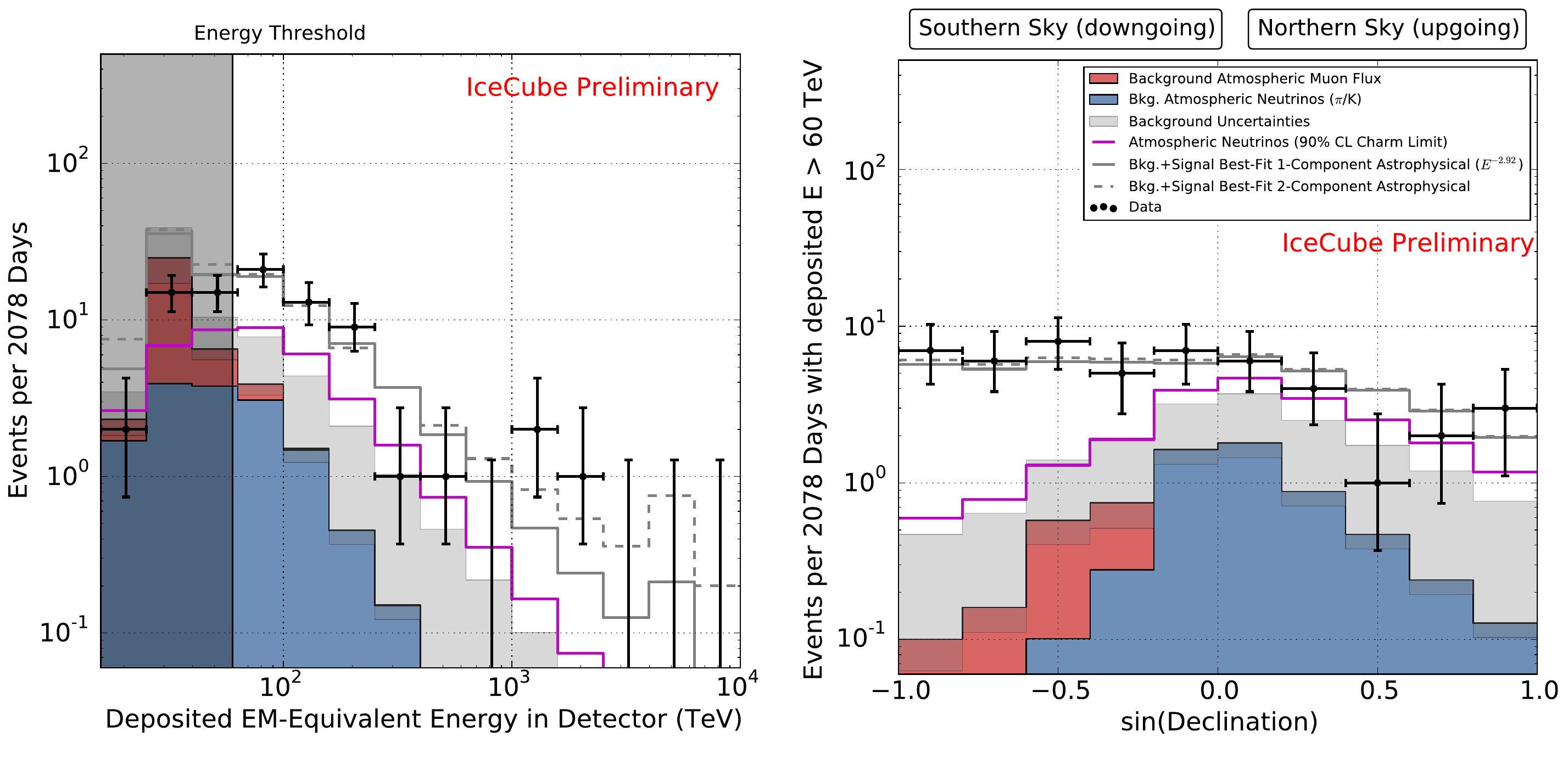}
\caption[]{{\bf Left Panel:} Deposited energies, by neutrinos interacting inside IceCube, observed in six years of data~\cite{Aartsen:2017mau}. The grey region shows uncertainties on the sum of all backgrounds. The atmospheric muon flux (blue) and its uncertainty is computed from simulation to overcome statistical limitations in our background measurement and scaled to match the total measured background rate. The atmospheric neutrino flux is derived from previous measurements of both the $\pi, K$, and charm components of the atmospheric spectrum \protect\cite{Aartsen:2013vca}. Also shown are two fits to the spectrum, assuming a simple power-law (solid gray) and a broken power-law (dashed gray). {\bf Right Panel:} The same data and models, but now showing the distribution of events with deposited energy above 60~TeV in declination. At the South Pole, the declination angle $\delta$ is equivalent to the distribution in zenith angle $\theta$ related by the identity, $\delta = \theta-\pi/2$. It is clearly visible that the data is flat in the Southern Hemisphere, as expected from the contribution of an isotropic astrophysical flux.}
\label{hese_energy}
\end{figure}

The starting event samples revealed the first evidence for neutrinos of cosmic origin~\cite{Aartsen:2013bka,Aartsen:2013jdh}. Events with PeV energies and with no trace of coincident muons that reveal either the decay products of a parent meson or an accompanying atmospheric shower are highly unlikely to be of atmospheric origin. The present seven-year data set contains a total of 60 neutrino events with deposited energies ranging from 60\,TeV to 10\,PeV that are likely to be of cosmic origin. The deposited energy and zenith dependence of the high-energy starting events~\cite{Aartsen:2017mau,Abbasi_2021} is compared to the atmospheric background in Fig.~\ref{hese_energy}. A purely atmospheric explanation of the observation is excluded at $8\sigma$.

\begin{figure*}[ht!]
\centering
\includegraphics[width=0.9\linewidth]{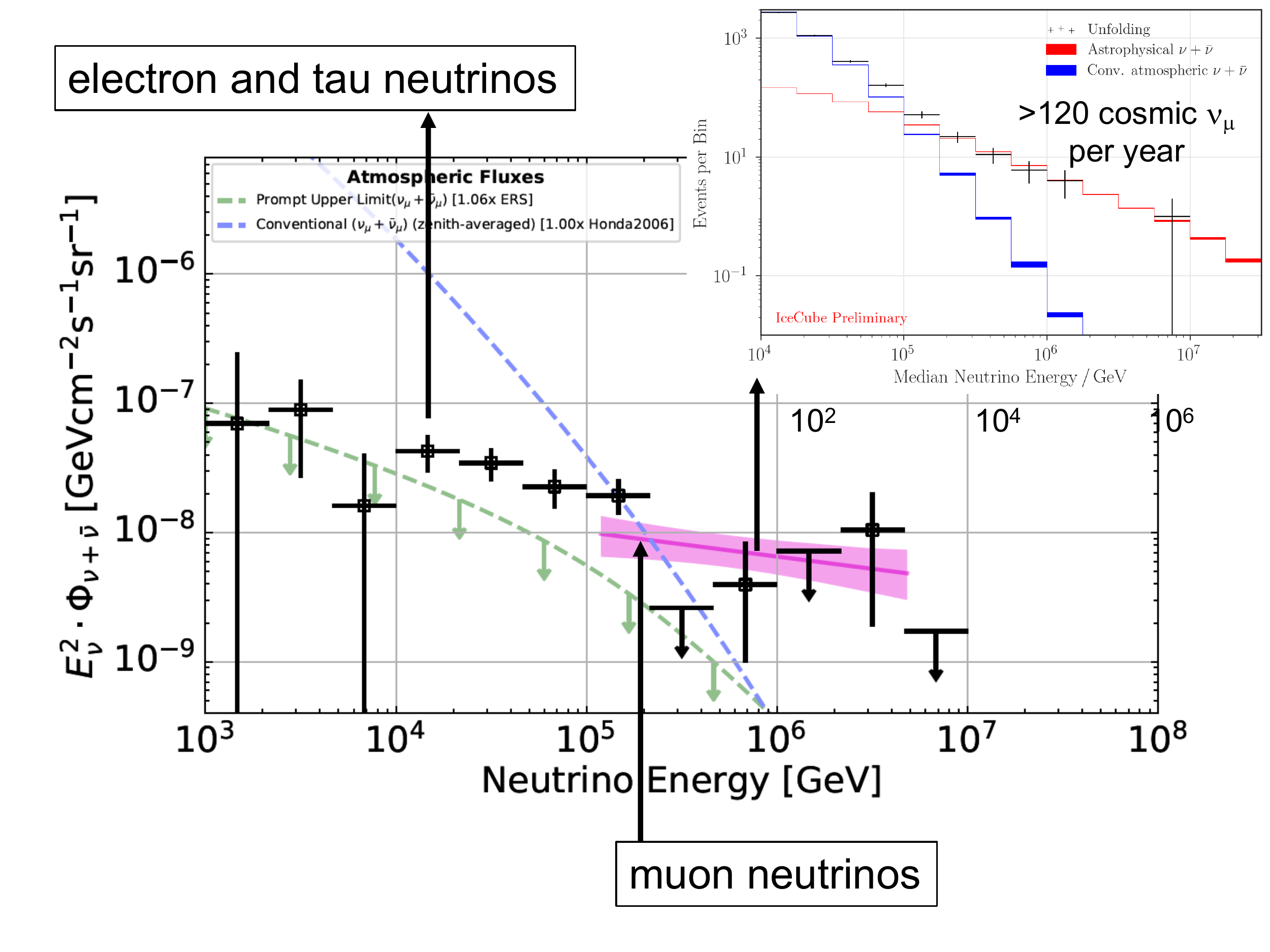}
\caption{The flux of cosmic muon neutrinos~\cite{Aartsen:2017mau} inferred from the eight-year upgoing-muon track analysis (red solid line) with $1\sigma$ uncertainty range (shaded range; from fit shown in upper-right inset) is compared with the flux of showers initiated by electron and tau neutrinos~\cite{IceCube:2020acn}. The measurements are consistent assuming that each neutrino flavor contributes an identical flux to the diffuse spectrum.}
\label{fig:showerstracks-2}
\end{figure*}

The flux of cosmic neutrinos has by now also been characterized with a range of other methods. Their results agree, pointing at extragalactic sources whose flux has equilibrated in the three neutrino flavors after propagation over cosmic distances~\cite{Aartsen:2015ivb}, with $\nu_e:\nu_\mu:\nu_\tau \sim 1:1:1$. Fig.~\ref{fig:showerstracks-2} shows the measurement of the cosmic neutrino flux of a recent analysis that specializes to showers only. These have been isolated from the atmospheric background down to energies below 10\,TeV~\cite{IceCube:2020acn}. The energy spectrum of $E^{-2.5}$ agrees with the measurement using upgoing muons with a spectral index of $E^{-2.4}$ above an energy of $\sim 100$\,TeV~\cite{Aartsen:2017mau}. In general, analyses reaching lower energies exhibit larger spectral indices with the updated 7.5 years starting-event sample~\cite{Abbasi_2021}, yielding a spectral index value of $-2.87\pm0.2$ for the 68.3\% confidence interval.

\subsection{Cosmic Tau Neutrinos}

\begin{figure}[ht!]
\centering
\includegraphics[width=0.9\linewidth]{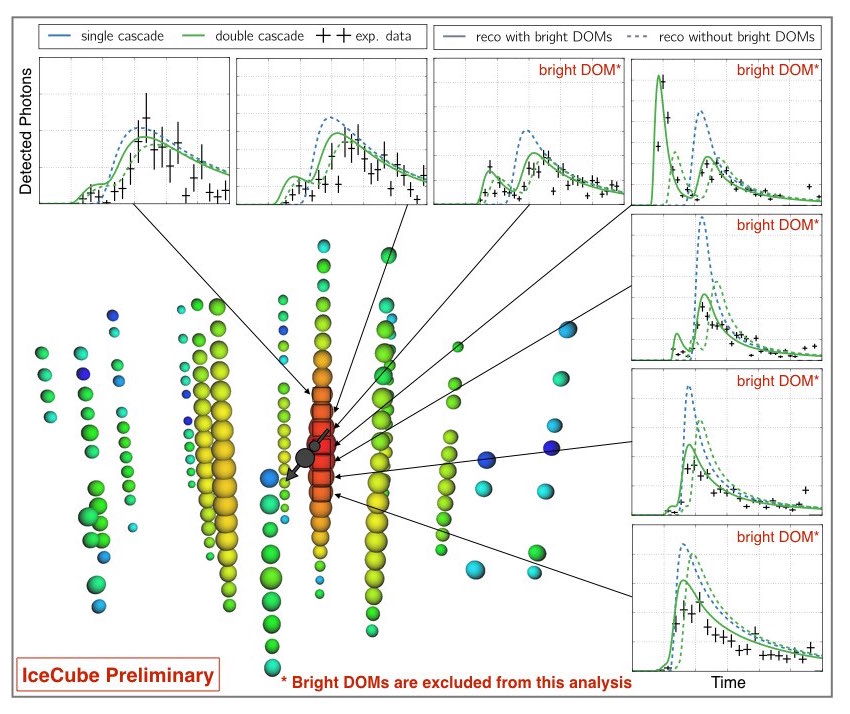}
\caption[]{Event view of a tau neutrino~\cite{Abbasi:2020zmr}. The Cherenkov photons associated with the production and subsequent decay of the tau neutrino are identified by the double-peaked photon count as a function of time for the bright DOMs, for instance, the one shown in the top-right corner. The best fit (solid line) corresponds to a 17\,m decay length and is far superior to fits assuming a single electromagnetic or hadronic shower (dashed lines).}
\label{fig:double_bang}
\end{figure}

We should comment at this point that there is yet another method to conclusively identify neutrinos that are of cosmic origin: the observation of high-energy tau neutrinos. Tau neutrinos are produced in the atmosphere by the oscillations of muon neutrinos into tau neutrinos, but only for neutrino energies well below 100\,GeV. Above that energy a tau neutrino must be of cosmic origin, produced in cosmic accelerators with a neutrino flux with a tau fraction of approximately one third. Tau neutrinos produce two spatially separated showers in the detector, one from the interaction of the tau neutrino and the second from the decay of the secondary tau produced~\cite{Learned:1994wg}. The mean decay length of the tau lepton is $\lambda_{\tau} = (E_\tau/m)\, c \,\tau \approx 50~{\rm m} \times (E_\tau/\rm{PeV})$, where $m$, $\tau$, and $E_\tau$ are the mass, lifetime, and energy of the tau, respectively. Two such candidate events have been identified~\cite{Abbasi:2020zmr}. An event with a decay length of 17\,m and a probability of 98\% of being produced by a tau neutrino is shown in Fig.~\ref{fig:double_bang}. The energies of the two showers are 9\,TeV and 80\,TeV. 

\begin{figure}[ht!]
\centering
\includegraphics[width=0.9\linewidth, trim=0 0 0 150, clip=true]{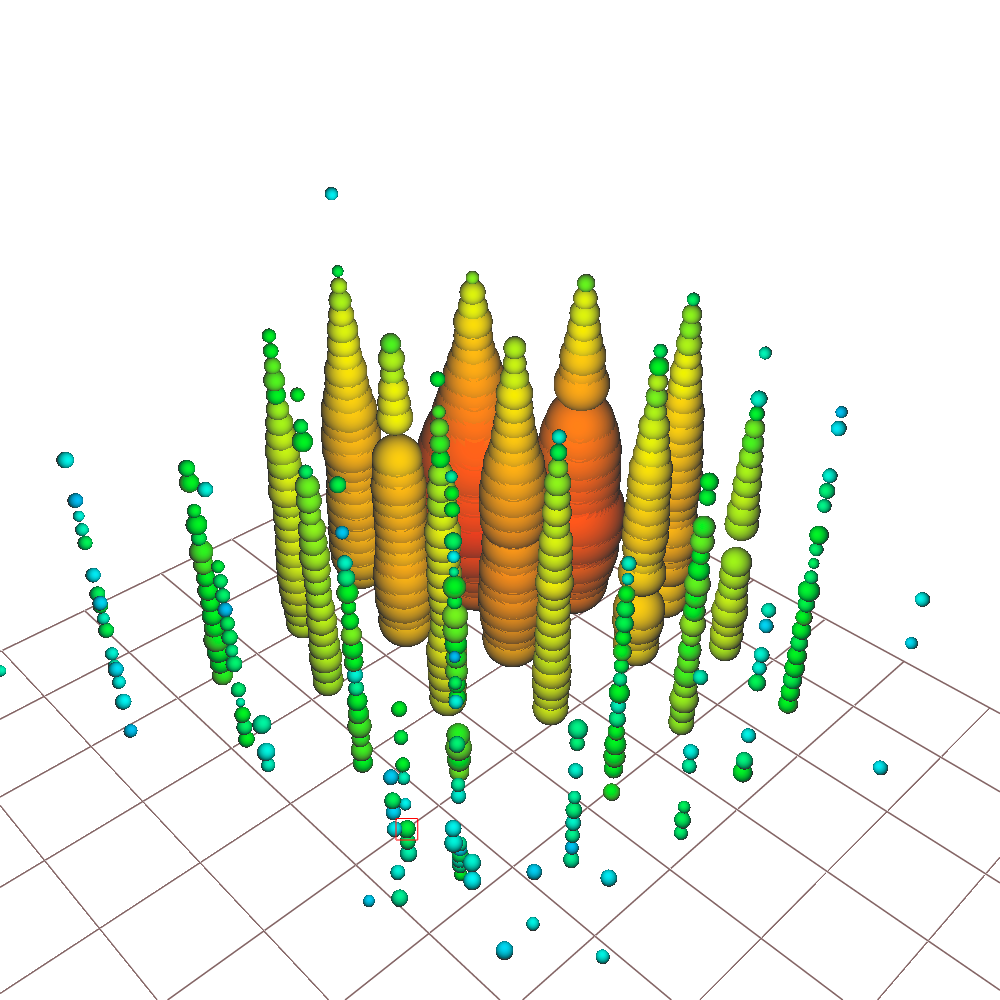}
\caption[]{Particle shower created by the Glashow resonance~\cite{IceCube:2021rpz}. Its energy is reconstructed at the resonant energy for the production of a weak intermediate boson $W^-$ in the interaction of an antielectron neutrino with an atomic electron in the ice. The properties of the secondary muons produced in the particle shower are consistent with the hadronic decay of a $W^-$ boson.}
\label{fig:glashow}
\end{figure}

Yet another independent confirmation of the observation of neutrinos of cosmic origin appeared in the form of the Glashow resonance event shown in Fig.~\ref{fig:glashow}. The event was identified in a dedicated search for partially contained showers of very high energy~\cite{IceCube:2021rpz}. The reconstructed energy of the shower is 6.3\,PeV, which matches the laboratory energy for the production of a weak intermediate $W^-$ in the resonant interaction of an electron antineutrino with an atomic electron: $\bar{\nu}_e + e^- \rightarrow W^- \rightarrow q + \bar{q}$. Given its high energy, the initial neutrino is cosmic in origin; it represents an independent discovery of cosmic neutrinos at the level of $5\sigma$. Assuming the Standard Model cross section, we expect 1.55 events in the data sample searched, assuming an antineutrino:neutrino ratio of 1:1 characteristic of a cosmic source producing an equal number of pions of all three electric charges. Taking into account the detector's energy resolution, the probability that the event is produced off resonance by deep inelastic scattering is only 0.01, assuming a spectrum with a spectral index of $\gamma = -2.5$. Furthermore, the presence of both muons and an electromagnetic shower is consistent with the hadronic decay of a $W^-$ produced on the Glashow resonance~\cite{IceCube:2021rpz}.

The observation of a Glashow resonance event indicates the presence of electron antineutrinos in the cosmic neutrino flux. Its unique signature provides a method to disentangle neutrinos from antineutrinos; their ratio distinguishes accelerators that produce neutrinos via $pp$ and $p\gamma$ interactions and is also sensitive to their magnetic field~\cite{IceCube:2021rpz}.

Finally, data from the ANTARES experiment are consistent with the observation of a flux of cosmic origin although with limited statistical significance~\cite{Albert_2018}.

\section{Multimessenger Astronomy}
\label{sec:multimessenger}

The most important message emerging from the IceCube measurements may not be apparent yet: the prominent role of neutrinos relative to photons in the extreme universe. Photons are inevitably produced in association with neutrinos when accelerated cosmic rays produce both neutral and charged pions in interactions with target material in the vicinity of the accelerator. While neutral pions decay into two gamma rays, $\pi^0\to\gamma+\gamma$, the charged pions decay into three high-energy neutrinos ($\nu$) and antineutrinos ($\bar\nu$) via the decay chain $\pi^+\to\mu^++\nu_\mu$ followed by $\mu^+\to e^++\bar\nu_\mu +\nu_e$ and the charged-conjugate process. On average, the four final state leptons equally share the energy of the charged pion. With these approximations, gamma rays and neutrinos carry on average 1/2 and 1/4 of the energy of the parent pion.

\begin{figure}[ht!]
\centering
\includegraphics[width=\columnwidth]{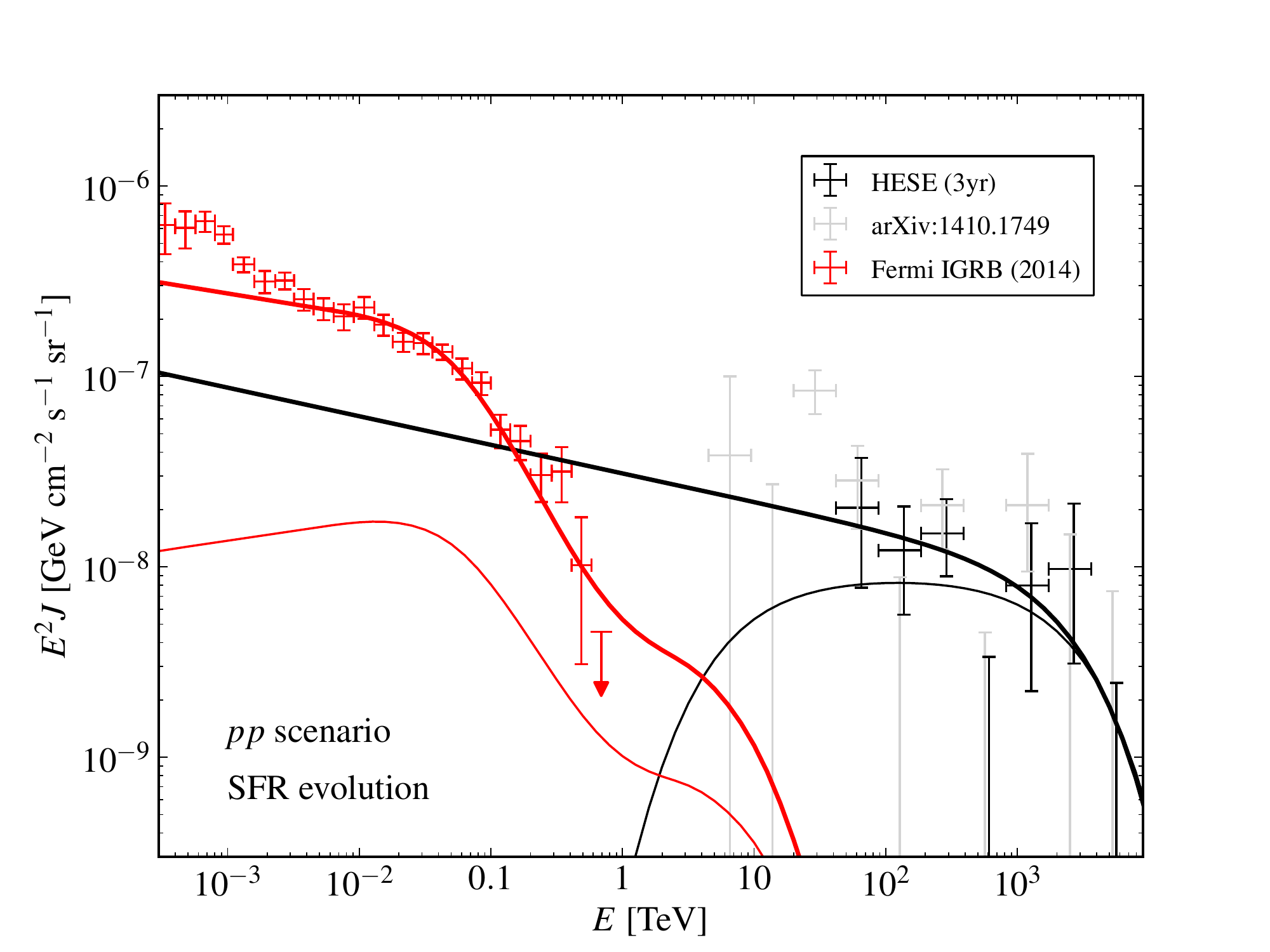}
\caption[]{A calculation illustrating that the photon flux that accompanies the neutrino flux (thick black line) measured by IceCube matches the gamma-ray flux (thick red line) observed by Fermi. We assume a $E^{-1.5}$ energy spectrum, star-formation redshift evolution and, importantly, gamma-ray transparent sources, i.e., pionic photons lose energy in the EBL but not in their sources. The black data points are early IceCube measurements~\cite{Aartsen:2014gkd,Aartsen:2014muf}. The result suggests that the decay products of neutral and charged pions from $pp$ interactions may be significant components of the nonthermal radiation in the extreme universe~\protect\cite{Murase:2013rfa}. (Introducing the cutoff on the high-energy flux, shown in the figure, does not affect the result.) The calculation is compared to an identical calculation adopting a spectral shape characteristic for the production of cosmic neutrinos on a gamma-ray target in the source (thin lines). While the pionic gamma-ray energy flux is now suppressed relative to the Fermi observations, the neutrino energy spectrum underestimates the IceCube observations and the conclusion that the sources are likely obscured is recovered after correct normalization to the most up to date measurements.~\cite{Murase:2013rfa}}
\label{fig:Fermi_vs_Icecube_gamma}
\end{figure}

Modeling of gamma ray observations generally indicated that the flux in pionic photons would be relatively small compared to a dominant flux of photons radiated by electrons in electromagnetic processes, synchrotron radiation, and inverse Compton scattering. These were generally assumed to dominate the energy production of sources in the Galaxy and beyond. IceCube observations show that this is not the case. To illustrate this point, we show in Fig.~\ref{fig:Fermi_vs_Icecube_gamma} the energy fluxes, $E^2 dN/dE$, of neutrinos and gamma rays in the universe. Clearly, the energy flux of cosmic neutrinos is comparable to the one for the highest energy gamma rays observed by the NASA Fermi satellite~\cite{Ackermann:2014usa}. This already indicates a more prominent role of hadronic processes than routinely anticipated. We will look at this match of the neutrino and gamma-ray energies in more detail.

The neutrino production rate $Q_{\nu_\alpha}$ (with typical units ${\rm GeV}^{-1} {\rm s}^{-1}$ and with the subscript $\alpha$ labeling the neutrino flavor) can be related to the one for charged pions $Q_\pi$ by
\begin{equation}
\sum_{\alpha}E_\nu Q_{\nu_\alpha}(E_\nu) \simeq 3\left[E_\pi Q_{\pi^\pm}(E_\pi)\right]_{E_\pi \simeq 4E_\nu}\,,
\label{eq:Qgamma}
\end{equation}
while, similarly, the production rate of pionic gamma rays is related to the one for neutral pions by
\begin{equation}
E_\gamma Q_{\gamma}(E_\gamma) \simeq 2\left[E_\pi Q_{\pi^0}(E_\pi)\right]_{E_\pi \simeq 2E_\gamma}\,.
\label{eq:PIONtoNU}
\end{equation}

Pion production in the interactions of cosmic rays with photon fields proceeds resonantly via the processes $p + \gamma \rightarrow \Delta^+ \rightarrow \pi^0 + p$ and $p + \gamma \rightarrow \Delta^+ \rightarrow \pi^+ + n$. These channels produce charged and neutral pions with probabilities 2/3 and 1/3, respectively. However, the additional contribution of nonresonant pion production changes this ratio to approximately 1/2 and 1/2. In contrast, cosmic rays interacting with matter produce equal numbers of pions of all three charges: $p+p \rightarrow n_\pi\,[\,\pi^{0}+\pi^{+} +\pi^{-}]+X$, where $n_\pi$ is the pion multiplicity. We thus obtain a charge ratio $K_\pi = n_{\pi^\pm}/n_{\pi^0} \simeq$\,2 and 1 for $pp$ and $p\gamma$ interactions, respectively. 

Eqs.~\ref{eq:Qgamma} and \ref{eq:PIONtoNU} can now be combined to obtain a direct relation between the gamma-ray and neutrino production rates:
\begin{equation}\label{eq:GAMMAtoNU}
\frac{1}{3}\sum_{\alpha}E^2_\nu Q_{\nu_\alpha}(E_\nu) \simeq \frac{K_\pi}{4}\left[E^2_\gamma Q_\gamma(E_\gamma)\right]_{E_\gamma = 2E_\nu}\,,
\end{equation}
where the factor $1/4$ accounts for the fact that two gamma rays are produced in the neutral pion decay with twice the energy of the accompanying neutrino, $\langle E_\nu\rangle/\langle E_\gamma\rangle\simeq 1/2$. Note that the relative production rate of gamma rays and neutrinos only depends on the ratio of charged-to-neutral pions produced without any reference to the cosmic-ray beam that initiates their production in the target. This powerful relation follows from the fact that pion production conserves isospin and nothing else.

Before applying this relation to data, one must recall that the universe is transparent to extragalactic neutrinos but not to the accompanying gamma rays. These will interact with microwave photons and other components of the extragalactic background light (EBL) to initiate an electromagnetic cascade that reaches Earth in the form of multiple photons of lower energy. The electromagnetic shower subdivides the initial PeV photon energy into multiple photons with GeV to TeV energies by the time it reaches Earth~\cite{Protheroe1993,Ahlers:2010fw}. If the source itself is opaque to gamma rays, the high-energy gamma rays will lose energy even before reaching the EBL to possibly emerge at Earth below the threshold of Fermi, at MeV energies and below.

In order to underscore the power of the multimessenger connection between photons and neutrinos, we first calculate the gamma-ray flux accompanying the diffuse cosmic neutrino flux observed by IceCube, which we describe by a power law with spectral index of $-2.15$, consistent with the neutrino data above an energy of 100~TeV. The result is shown in Fig.~\ref{fig:Fermi_vs_Icecube_gamma} assuming gamma-ray transparent sources and equal multiplicities of all three pion charges, i.e., $K_\pi=2$. The cascaded gamma-ray energy flux resulting from the pionic photons accompanying the neutrino flux matches the energy flux of extragalactic gamma rays measured by Fermi.

Clearly, in this illustration, the slope and overall normalization of the neutrino spectrum have been adjusted to not exceed the isotropic extragalactic gamma-ray background observed by the Fermi satellite. We conclude that the high-energy cosmic neutrino flux above 100~TeV shown in Fig.~\ref{fig:Fermi_vs_Icecube_gamma} saturates the Fermi measurement for the highest photon energies; higher normalization and larger spectral index of the neutrino flux will result in a gamma-ray flux that exceeds the Fermi observations. Fitting the IceCube data with a $E^{-2.5}$ spectral index, closer to the present observations, results in larger neutrino energy fluxes at energies below 100\,TeV for both neutrinos and their accompanying photons. After cascading in the EBL, the latter exceeds the Fermi observations. There is no conflict here; in this case, we conclude that the assumption that the sources themselves are transparent to photons is untenable. The resolution is as mentioned above, that the photons lose energy in the source even before entering the EBL and, as a result, reach Earth with energies that are below the detection threshold of the Fermi satellite, at MeV energy or below.

Alternatively, the target for producing the neutrinos may be photons. This changes the value of $K_\pi$ and, more importantly, the shape of the energy spectrum; for a detailed discussion, see, for instance, Ref.~\citen{Yoshida_2020}. Yielding an energy spectrum that peaks near PeV energies, as shown in Fig.~\ref{fig:Fermi_vs_Icecube_gamma}, the contribution to the Fermi flux is suppressed at lower energies relative to the power law assumed in Fig.~\ref{fig:Fermi_vs_Icecube_gamma}. However, as was the case for $pp$ interactions, fits that do not exceed the Fermi data tend not to accommodate the cosmic neutrino spectrum below 100\,TeV. This is illustrated in Fig.~\ref{fig:Fermi_vs_Icecube_gamma} where we obtain a neutrino spectrum well below the Fermi observations at the price of not fitting the overall normalization of the neutrino data. If sources of the TeV-PeV neutrinos are transparent to gamma rays with respect to two-photon annihilation, tensions with the isotropic diffuse gamma-ray background measured by Fermi seem unavoidable, independently of the production mechanism~\cite{Murase:2013rfa}.

The conclusion is inescapable that the energy fluxes of neutrinos and gamma rays in the extreme universe are qualitatively the same. Furthermore, the IceCube observations point at contributions to the diffuse flux from gamma-obscured sources.

We therefore anticipate that multimessenger studies of gamma-ray and neutrino data will be a powerful tool to identify and study the cosmic ray accelerators that produce cosmic neutrinos. Accordingly, IceCube has developed methods, most promising among them the real-time multiwavelength observations with astronomical telescopes, to identify the sources and build on the discovery of cosmic neutrinos to launch a new era in astronomy~\cite{Aartsen:2016qbu,Aartsen:2016lmt}.

An important lesson for multiwavelength astronomy is that strong high-energy gamma-ray emitters may not be the best candidate neutrino sources. For instance, IceCube does not observe neutrinos from gamma-ray bursts, but this result covers a sample of bursts that are strong gamma-ray emitters~\cite{IceCube:2017amx}. A handful of bursts have been identified that are obscured in gamma rays and may instead provide the key to detecting a neutrino signal. The strongest sources in the 10-year IceCube sky map, the galaxies NGC 1068 and TXS 0506+O56, are gamma-ray-obscured sources at least at the time that they emitted neutrinos. We will discuss this next.

Interestingly, the common energy density of photons and neutrinos is also comparable to that of the ultra-high-energy extragalactic cosmic rays~\cite{Aab:2015bza}.

\section{Identifying Neutrino Sources: the Supermassive Black Hole TXS 0506+056}\label{TXS}

Phenomenological studies~\cite{Ajello:2011zi,DiMauro:2013zfa} and recent data analyses~\cite{TheFermi-LAT:2015ykq,Zechlin:2015wdz,Lisanti:2016jub} have converged on the fact that Fermi's extragalactic gamma-ray flux shown in Fig.~\ref{fig:Fermi_vs_Icecube_gamma} is dominated by blazars, AGN with jets pointing at Earth. It is tempting to conclude, based on the matching energy fluxes of photons and neutrinos discussed in the previous section, that the unidentified neutrino sources contributing to the diffuse neutrino flux have already been observed as strong gamma-ray emitters. This is not the case. A dedicated IceCube study~\cite{Aartsen:2016lir} correlating the arrival directions of cosmic neutrinos with Fermi blazars shows no evidence of neutrino emission from these sources. The limit leaves room for a contribution of Fermi blazars to the diffuse cosmic neutrino flux below the 10\% level. Surprisingly, the multimessenger campaign launched by the neutrino alert IC-170922A~\cite{IceCube:2018dnn} identified the first source of cosmic neutrinos as a Fermi ``blazar." We will discuss how the multiwavelength data shed light on the apparent contradiction.

Since 2016, the IceCube multimessenger program has grown from issuing Galactic supernova alerts~\cite{IceCube:2011cwc} and matching neutrinos with early LIGO/Virgo gravitational wave candidates to a steadily expanding set of automatic filters that select in real time rare, very high energy neutrino events that are likely to be cosmic in origin~\cite{Abbasi:2020aae}. Within less than one minute of stopping in the instrumented Antarctic ice, the arrival directions of the neutrinos are reconstructed and automatically sent to the Gamma-ray Coordinate Network for potential follow-up by astronomical telescopes.

\subsection{Observation of a Cosmic Neutrino Source: TXS 0506+056}

\begin{figure}[ht!]
    \centering
    \includegraphics[width=0.9\textwidth]{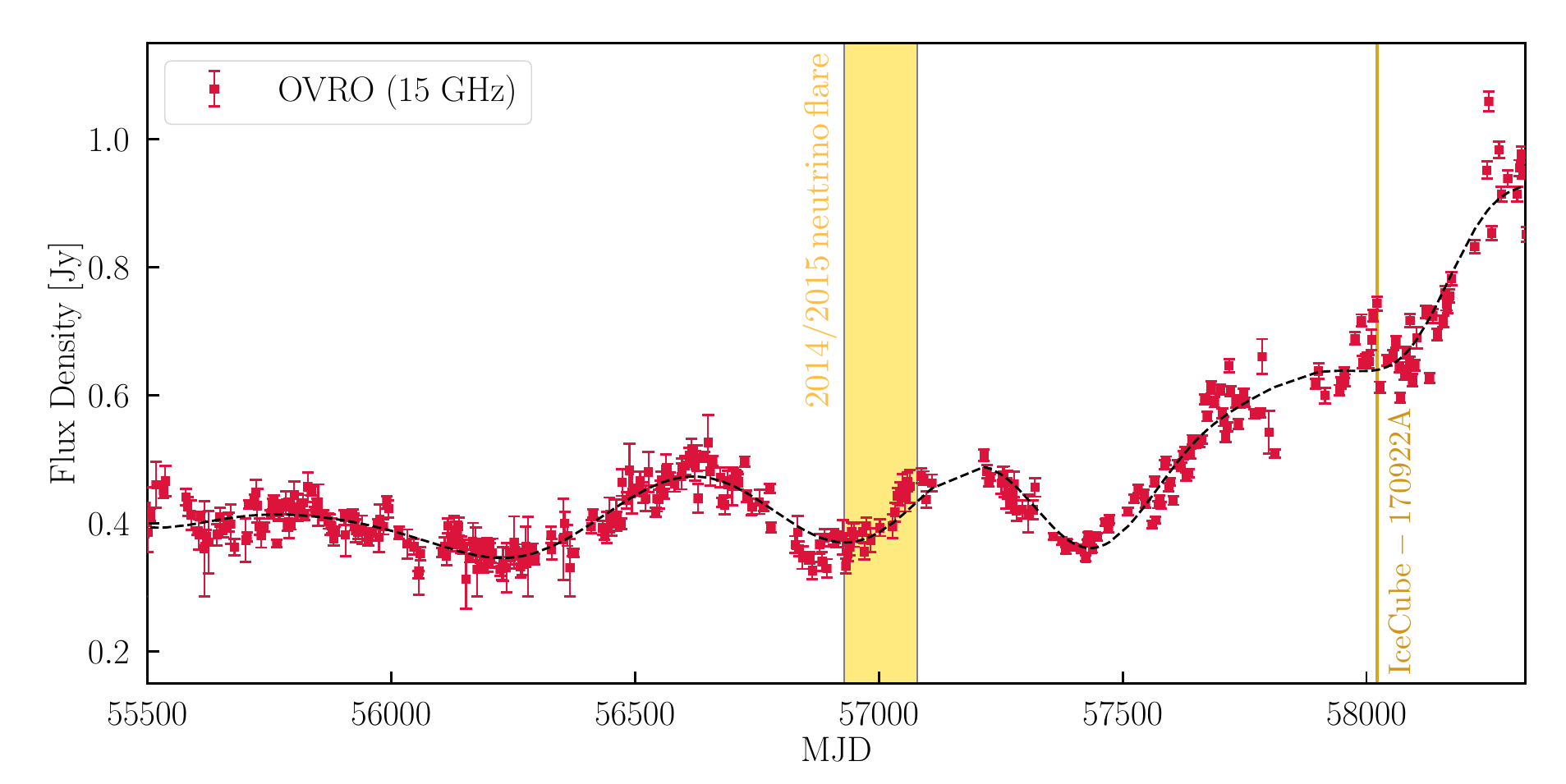}
    \caption{TXS 0506+056 radio light curve from Owen Valley Radio Observatory (OVRO) at 15 GHz (red). The dashed line illustrates the pattern of the radio flux density. The 2014/15 110-day neutrino flare (yellow band) and the IceCube-170922A episodes are shown.}
    \label{fig:txs-ovro}
\end{figure}

On September 22, 2017, the tenth such alert, IceCube-170922A~\cite{2017GCN.21916....1K}, reported a well-reconstructed muon that deposited 180 TeV inside the detector, corresponding to a most probable energy of the parent neutrino of 290\,TeV. Its arrival direction was aligned with the coordinates of a known Fermi blazar, TXS 0506+056, to within $0.06^\circ$. The source was ``flaring" with a gamma-ray flux that had increased by a factor of seven in recent months. A variety of estimates converged on a probability on the order of $10^{-3}$ that the coincidence was accidental. The identification of the neutrino with the source reached the level of evidence, but not more. What clinched the association was a series of subsequent observations, culminating with the optical observation of the source switching from an ``off" to an ``on" state two hours after the emission of IC-170922A, conclusively associating the neutrino with TXS 0506+056~\cite{2020ApJ...896L..19L}. The sequence of observations can be summarized as follows:
\begin{itemize}
    \item The redshift of the host galaxy, a known blazar, was measured to be $z\simeq0.34$~\cite{Paiano:2018qeq}. It is important to realize that nearby blazars like the Markarian sources are at a distance that is ten times closer, and therefore TXS 0506+056, with a similar flux despite the greater distance, is one of the most luminous sources in the universe. This suggests that it belongs to a special class of sources that accelerate proton beams in dense environments, revealed by the neutrino. That the source is special eliminates any conflict between its observation and the lack of correlation between the arrival directions of IceCube neutrinos and the bulk of the blazars observed by Fermi~\cite{Aartsen:2016lir}. Such limits implicitly assume that all sources in an astronomical category are identical, and this is a strong, unstated assumption as underscored by this observation.
    \item Originally detected by NASA's Swift~\cite{2017ATel10792....1E} and Fermi~\cite{2017ATel10791....1T} satellites, the alert was followed up by ground-based air Cherenkov telescopes~\cite{2017ATel10817....1M}. MAGIC detected the emission of gamma rays with energies exceeding 100 GeV starting several days after the observation of the neutrino~\cite{Ahnen:2018mvi}. Given its distance, this establishes the source as a relatively rare TeV blazar.
    \item Informed where to look, IceCube searched its archival neutrino data up to and including October 2017 for evidence of neutrino emission at the location of TXS 0506+056~\cite{IceCube:2018cha}. When searching the sky for point sources of neutrinos, two analyses have been routinely performed: one that searches for steady emission of neutrinos and one that searches for flares over a variety of timescales. Evidence was found for 19 high-energy neutrino events on a background of fewer than 6 in a burst lasting 110 days. This burst dominates the integrated flux from the source over the last 9.5 years of archival IceCube data, leaving the 2017 flare as a second subdominant feature. We note that this analysis applied a published prescription to data; the chance that this observation is a fluctuation is small.
    \item Radio interferometric images \cite{Britzen:2019fje,Kun_2018} of the source revealed a jet that loses its tight collimation beyond 5 milliarcseconds running into material or intense radiation fields that are likely to be the target for producing the neutrinos. The nature of the target is still a matter of debate. Speculations include the merger with another galaxy that may supply plenty of material to interact with the jet of the dominant galaxy. Alternatively, the jet may interact with the dense molecular clouds of a star-forming region or simply with supermassive stars in the central region of the host galaxy \cite{Britzen:2019fje,Kun_2018}. Also, in a so-called structured jet, the accelerated protons may catch up and collide with a slower moving and denser region of jetted photons. Additionally, the VLBA data reveal that the neutrino burst occurs at the peak of enhanced radio emission at 15 GHz, which started five years ago; see Fig.~\ref{fig:txs-ovro}. The radio flare may be a signature of a galaxy merger; correlations of radio bursts with the process of merging supermassive black holes have been anticipated~\cite{Gergely_2009}. 
    \item The MASTER robotic optical telescope network has been monitoring the source since 2005 and detected its strongest time variation in the last 15 years to occur two hours after the emission of IC170922, with a second variation following the 2014-15 burst~\cite{2020ApJ...896L..19L}. The blazar switches from the ``off" to the ``on" state two hours after the emission of the neutrino. After an episode of monitoring the uniformity of their observations of the source in the first quarter of 2020, they argue that the time variation detected on September 22, 2017 conclusively associates the source with the neutrino~\cite{2020ApJ...896L..19L}.
\end{itemize}

Additionally, it is important to note the fact that the high-energy photon and neutrino spectra covering the 2014 burst are consistent with a hard $E^{-2}$ spectrum, which is expected for a cosmic accelerator. In fact, the gamma-ray spectrum shows a hint of flattening beyond $E^{-2}$ during the 110-day period of the 2014 burst \cite{Padovani:2018acg,Garrappa:2018}.

In summary, both the multiwavelength campaign~\cite{IceCube:2018dnn} and the observation of an earlier burst of the same source in archival neutrino data provide statistically independent~\cite{IceCube:2018cha} evidence for the association of neutrinos with TXS 0506+056 at the 3 and $3.7\sigma$ level, respectively. The significances contributed by the optical and TeV associations on timescales of hours and days are not taken into account, and it is not straightforward to do so because of their a posteriori nature. We conclude, however, that the association of neutrinos with the source summarized above is compelling.

Other IceCube alerts have triggered intriguing observations. Following up on a July 31, 2016, neutrino alert, the AGILE collaboration, which operates an orbiting X-ray and gamma-ray telescope, reported a day-long blazar flare in the direction of the neutrino one day before the neutrino detection~\cite{Lucarelli:2017hhh}. A tentative but very intriguing association of an IceCube alert~\cite{Stein_2021} has been made with a tidal disruption event, an anticipated source of high-energy neutrino emission. Even before IceCube issued automatic alerts, in April 2016, the TANAMI collaboration argued for the association of the highest energy IceCube event at the time, dubbed ``Big Bird,'' with the flaring blazar PKS B1424-418~\cite{Kadler_2016}. Interestingly, the event was produced at a minimum of the Fermi flux~\cite{Kun_2021}, indicating a gamma-ray-obscured source. AMANDA, IceCube's predecessor, observed~\cite{Bernardini:2006az} three neutrinos in coincidence with a rare flare of the blazar 1ES 1959+650 detected by the Whipple telescope in 2002~\cite{Daniel:2005rv}. However, none of these identifications reach the significance of the observations triggered by IC-170922A.

\subsection{The Blueprint of the TXS 0506+056 Beam Dump?}

Blazars are accelerators with their jets pointing at Earth. The gamma-ray community has developed a routine procedure for modeling their spectrum with two components: a lower energy component produced by synchrotron radiation by the electron beam and a high-energy component resulting from the inverse Compton scattering of (possibly the same) photons by accelerated electrons; for a recent discussion, see Ref.~\citen{Biteau_2020}. Such a source cannot accommodate the TXS observations for two reasons: an electron beam does not produce pions that decay into neutrinos, and, even with the presence of protons in the beam, a target is required to produce the parent pions. It is also evident that a source that emits high-energy gamma rays is transparent to $\gamma\gamma$ absorption and unlikely to host the target material to produce neutrinos. The opacity for $\gamma\gamma$ interactions to absorb photons is typically two orders of magnitude larger than the one for $p\gamma$ interactions to produce pions and neutrinos~\cite{Halzen:2018iak}.

With the opacity to photons about two orders of magnitude larger, the target producing the neutrinos is unlikely to be transparent to TeV photons, only to photons with tens of GeV energy, and that is indeed what is observed by Fermi at the time of the 2014 flare. There is also evidence that, temporarily, TXS 0506+056 was a gamma-ray-obscured source at the time the IC-170922A neutrino was emitted. The optical observations show a dramatic transition of the blazar from the ``off" to the ``on" state two hours after the emission of the neutrino, resulting additionally in the doubling of its total optical luminosity~\cite{2020ApJ...896L..19L}. Also at the time of IC-170922 the ground-based atmospheric gamma-ray telescopes observed rapid variations in the flux around the time of the neutrino emission~\cite{IceCube:2018dnn}, with the gamma-ray emission observed by MAGIC only emerging after several days~\cite{Ahnen:2018mvi}.

\begin{figure}[ht!]\centering
\includegraphics[width=0.60\columnwidth,angle=-90]{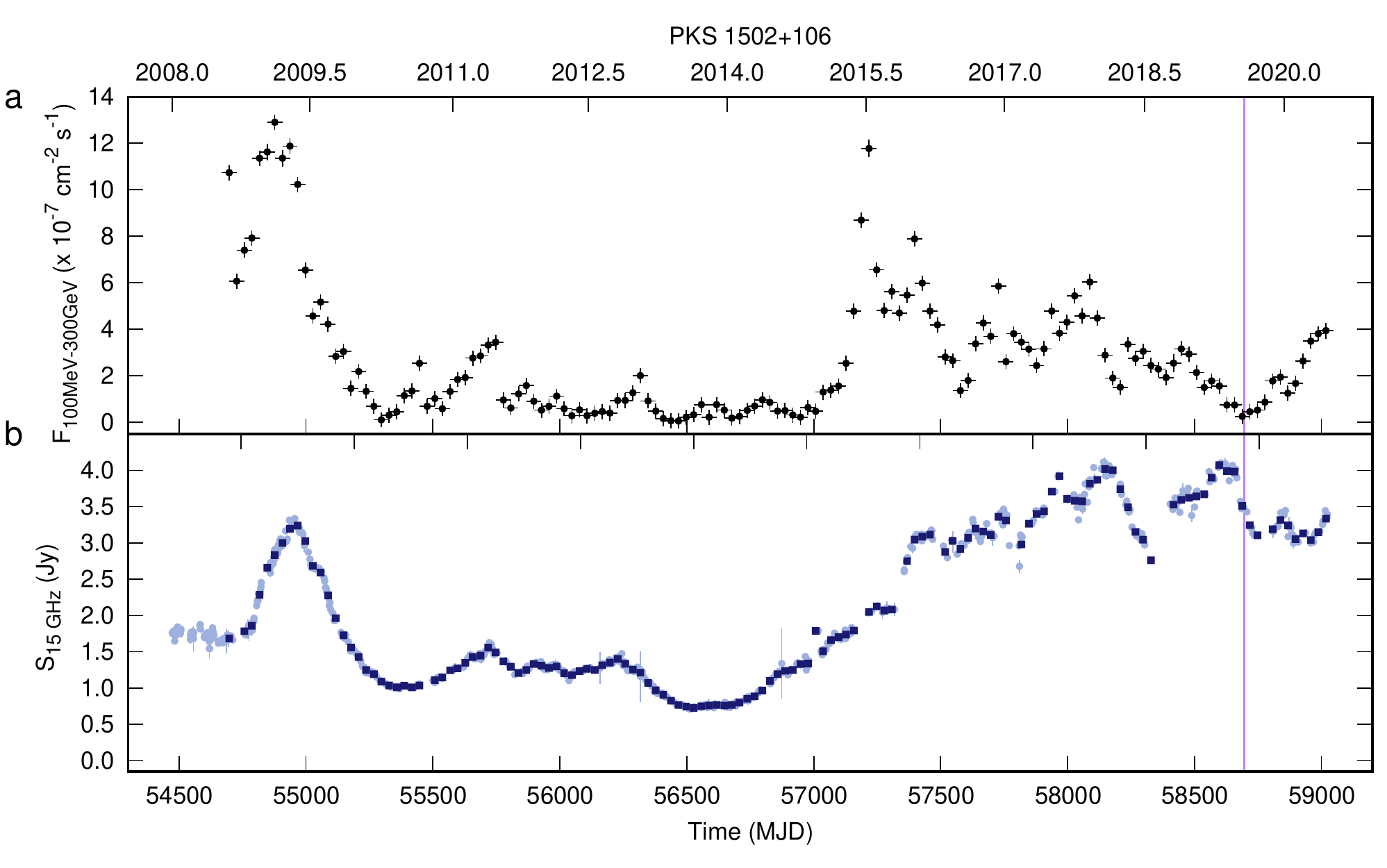}
\caption[]{Temporal variation of the $\gamma$-ray and radio brightness of PKS\,1502+106~\cite{Kun_2021}. {\bf Top Panel:} Fermi-LAT likelihood light curve integrated between 100\,MeV and 300\,GeV (marked by black dots with error bars). {\bf Bottom Panel:} OVRO flux density curve of PKS\,1502+106 plotted with light blue dots, which is superimposed by the radio flux density curve binned to the Fermi-LAT light curve (marked with dark blue squares). The detection time of the neutrino IC-190730A is labeled by a vertical purple line.}
\label{fig:fermiovro}
\end{figure}

A more direct indication that strong neutrino emitters are gamma-ray-obscured is indicated by a more recent alert, IC-190730A, sent by IceCube on July 30, 2019. A well-reconstructed 300-TeV muon neutrino was observed in spatial coincidence with the blazar PKS 1502+106~\cite{2019ATel12971....1L}. With a reconstructed energy just exceeding that of IC-170922A, it is the highest energy neutrino alert so far. OVRO radio observations~\cite{2019ATel12996....1K} show that the neutrino is coincident with the peak flux density of a flare at 15 GHz that started five years prior~\cite{2016A&A...586A..60K}, matching the similar long-term radio outburst of TXS 0506+056 at the time of IC-170922A; see Fig.~\ref{fig:txs-ovro}. Even more intriguing is the fact that the gamma-ray flux observed by Fermi shows a clear minimum at the time that the neutrino is emitted; see Fig.~\ref{fig:fermiovro}. We infer that at this time the jet meets the target that produces the neutrino. Inevitably, the accompanying high-energy gamma rays will be absorbed and the bulk of their electromagnetic energy will cascade down to energies below the Fermi threshold, i.e., MeV or below. Accumulating evidence indicates  that  cosmic  neutrinos  are  produced  by  temporarily  gamma-suppressed blazars or some other category of AGN.

The coincidence of the two highest energy alerts to date with extended periods of radio emission has led to the speculation that all~\cite{Plavin:2020mkf}, or at least some~\cite{Hovatta:2020lor}, cosmic neutrinos originate in radio-bright galaxies. The significance of the correlation and its physical origin, if any~\cite{Zhou:2021rhl}, are debated.

We suggest that neutrino sources are special, with properties unlikely to match any astronomical classification; astronomical observations catalog accelerators, not neutrino sources. If the neutrino flux observed from TXS 0506+056 over the last decade were typical of all blazars, these would overproduce the total diffuse flux observed by IceCube by well over an order of magnitude\cite{Halzen:2018iak}. The diffuse neutrino flux limits the density of such sources in the universe to $\sim 5 \times 10^{-10} \rm\,per\,Mpc^3$, i.e., about 5\% of all blazars, and more likely an even smaller fraction of all AGN~\cite{Halzen:2019qkf}. Attempts by IceCube and others to find a correlation between the directions of high-energy neutrinos and $\it{all}$ Fermi blazars must inevitably be unsuccessful, as was the case in Ref.~\citen{Aartsen:2016lir}.

The nature of this special class of sources has not been settled. One straightforward explanation could be that a subclass of AGN, selected by redshift evolution, are powerful proton accelerators producing neutrinos in the past and no longer active today. This assumption accommodates the relatively large redshift of TXS 0506+056, which would be the closest among a set of sources that only accelerated cosmic rays at early redshifts~\cite{Neronov_2020}.

Alternatively, in merging galaxies there is plenty of material for accelerated cosmic rays to interact with the jet of the dominant galaxy. Merger activity in active galaxies in not uncommon. The fresh material provides optically thick environments and allows for rapid variation of the Lorentz factors. A cursory review of the literature on the production of neutrinos in galaxy mergers is sufficient to conclude that they can indeed accommodate the observations of both the individual sources discussed above and the total flux of cosmic neutrinos~\cite{Kashiyama:2014rza,Yuan:2017dle,Yuan:2018erh}. Besides mergers, some form of structured jet where the accelerated protons collide with a slower moving and denser region of jetted photons is a possibility. The jet could also interact with dense molecular clouds of a star-forming region or simply with supermassive stars in the central region of the host galaxy~\cite{Britzen:2019fje,Kun_2018}.

The observation that the energy flux in neutrinos and very high energy cosmic rays are similar supports the fact that cosmic rays must be highly efficient at producing neutrinos, requiring a large target density that renders them opaque to high-energy gamma rays. A consistent picture emerges with the source opacity $\tau_{p\gamma}$ exceeding a value of 0.8~\cite{Halzen:2018iak,Halzen:2019qkf}, resulting in a gamma-ray cascade where photons lose energy in the source before cascading to yet lower energies in the extragalactic background light. Some of their energy emerges below the Fermi threshold by the time they reach Earth. This is consistent with the discussion in the previous section that the multimessenger relation between neutrinos and gamma rays points at obscured sources.

We previously mentioned the evidence emerging from 10 years of IceCube data that the arrival directions of cosmic neutrinos are no longer isotropic~\cite{Carver:2019jcd}. The anisotropy results from four sources—TXS 0506+056 among them—that show evidence for clustering above the 4$\sigma$ level (pretrial). The strongest of these sources is the nearby active galaxy NGC 1068. There is evidence for shocks near the core and for molecular clouds with column density reaching $\sim 10^{25} \rm \,  cm^{- 2}$\cite{Marinucci:2015fqo}. Similar to TSX 0506+056, a merger onto the black hole is observed—either with a satellite galaxy or, more likely, with a star-forming region~\cite{2014A&A...567A.125G} accounting for the molecular clouds. This major accretion event may be the origin of the increased neutrino emission.

Although it is obviously challenging to provide a final conclusion on the origin of neutrinos and cosmic rays, we should not lose sight of the fact that high-energy neutrino astronomy exists and that IceCube has demonstrated it has the tools to reveal the extreme universe with more data, or, more realistically, with a larger detector.

\section{From Discovery to Astronomy: Larger Telescopes with Better Angular Resolution}

Neutrino astronomy has reached a stage that reminds us of the time when Trevor Weekes's 11-meter ground-based Cherenkov telescope had established one convincing TeV source: the Crab. Even today we do not know how the Crab produces gamma rays that track the pulsations of the pulsar at TeV energy. We have to take the lead on TeV astronomy: building more and better telescopes.

Following the pioneering work of DUMAND~\cite{Babson:1989yy}, several neutrino telescope projects were initiated in the Mediterranean Sea and in Lake Baikal in the 1990s~\cite{BAIKAL:1997iok,Aggouras:2005bg,Aguilar:2006rm,Migneco:2008zz}. In 2008, the construction of the ANTARES detector off the coast of France was completed. It demonstrated the feasibility of neutrino detection in the deep sea and has provided a wealth of technical experience and design solutions for deep-sea components. An international collaboration has started construction of a multi-cubic-kilometer neutrino telescope in the Mediterranean Sea, KM3NeT~\cite{Adrian-Martinez:2016fdl}. KM3NeT in its second phase~\cite{Adrian-Martinez:2016fdl} will consist of two units for astrophysical neutrino observations, each consisting of 115 strings carrying more than 2,000 optical modules. Since April 2021, six are operational and taking data~\cite{sinopoulou2021atmospheric}.

A parallel effort is underway in Lake Baikal with the construction of the deep underwater neutrino telescope Baikal-GVD (Gigaton Volume Detector)~\cite{Avrorin:2015wba}. The first GVD cluster was upgraded in the spring of 2016 to its final size: 288 optical modules, a geometry of 120 meters in diameter and 525 meters high, and an instrumented volume of 6 Mton. Each of the eight strings consists of three sections with 12 optical modules. At this time, 7 of the 14 clusters have been deployed, reaching a sensitivity close to the diffuse cosmic neutrino flux observed by IceCube.

IceCube itself is deploying seven new strings at the bottom of the detector array that have been designed as an incremental extension of the DeepCore detector and as a test bed for the technologies of a next-generation detector. The new instrumentation will dramatically boost IceCube’s performance at the lowest energies, increasing the samples of atmospheric neutrinos by a factor of ten. New calibration devices will advance our understanding of the response of the light sensors in both current and new strings, resulting in improved reconstructions of cascade events, better identification of tau neutrinos, and an enhanced pointing resolution of muon neutrinos that could approach the 0.1 degree level for the highest energy events of cosmic origin. The improved calibration of the existing sensors will also enable a reanalysis of more than ten years of archival data and significantly increase the discovery potential for neutrino sources before the construction of a second-generation instrument.

Further progress requires a larger instrument. Therefore, as a next step, IceCube proposes to instrument $10\rm\,km^3$ of glacial ice at the South Pole, capitalizing on the large absorption length of light in ice to thereby increase IceCube's sensitive volume by an order of magnitude~\cite{IceCube-Gen2:2020qha}. This large gain is made possible by the unique optical properties of the Antarctic glacier revealed by the construction of IceCube. Exploiting the extremely long photon absorption lengths in the deep Antarctic ice, the spacing between strings of light sensors will be increased from 125 to close to 250 meters without significant loss of performance of the instrument at TeV energies and above. The instrumented volume can therefore grow by one order of magnitude while keeping the instrumentation and its budget at the level of the current IceCube detector. The new facility will increase the rates of cosmic events from hundreds to thousands over several years. The superior angular resolution of the longer muon tracks will allow for the discovery of cosmic neutrino sources, currently seen at the $\sim 3 \sigma$ level in the 10-year sky map; see Fig.~\ref{fig:10year}.

\section{Acknowledgements}

Discussion with collaborators inside and outside the IceCube Collaboration, too many to be listed, have greatly shaped this presentation. Thanks. I would like to single out Markus Ahlers, Qinrui Liu and Ali Kheirandish for contributing to aspects of this manuscript.
This research was supported in part by the U.S. National Science Foundation under grants~PLR-1600823 and PHY-1913607 and by the University of Wisconsin Research Committee with funds granted by the Wisconsin Alumni Research Foundation.

\bibliographystyle{ws-ijmpd}
\bibliography{bib_new}
\end{document}